\documentclass[11pt,a4paper]{article}

\usepackage{amssymb}
\usepackage{graphicx}
\usepackage{bm}
\usepackage{multicol}
\usepackage{xcolor}

\usepackage{mathrsfs}
\usepackage{cite}
\usepackage{float}
\begin{document}

\title{Perturbative  QCD fitting of  KEDR and BESIII  $e^+e^-$ data  for R(s) and $\alpha_s$ determination}

\author{
A. L. Kataev${}^{a,b}$\footnote{e-mail:kataev@ms2.inr.ac.ru}~~and K. Yu. Todyshev ${}^{c,d}$\footnote{e-mail:todyshev@inp.nsk.su}\\
\\
${}^a${\small{\em Institute for Nuclear Research of the Russian Academy of Sciences,}}\\
{\small {\em 117312, Moscow, Russia}}\\
${}^b${\small{\em Bogoliubov Laboratory of Theoretical Physics, Joint Institute for Nuclear Research}}\\
{\small {\em 141980, Dubna, Russia}}\\
${}^c${\small{\em Budker Institute of Nuclear Physics of the Russian
    Academy of Sciences,}}
{\small{\em 630090, Novosibirsk, Russia
}}\\
${}^d${\small{\em Novosibirsk State University, 630090, Novosibirsk, Russia}} \\
}

\maketitle

\begin{abstract}
         The experimental  data collected by      KEDR and BESIII  collaborations
                 at the energies  below charm  quark  thresholds  are compared with the massless
                 QCD   expressions for the $e^+e^-$ annihilation  R-ratio
                  truncated at different orders of   perturbation  theory.
                The fits demonstrate  the dependence  of  the extracted $\alpha_s(M_Z)$ values on the
                 orders of  truncation of the corresponding  approximations.  The next-to-leading order, next-to-next-to-leading order
                 and next-to-next-to-next-to-leading order  fits
                 of the combined  KEDR  data and BESIII  data ,
                 truncated at  the  scale of mass of  $J/\Psi$ meson,
                 give the following results $\alpha_s(M_Z)=0.1151_{-0.0069}^{+0.0052}$,
                $\alpha_s(M_Z)=0.1190_{-0.0081}^{+0.0064}$and
                $\alpha_s(M_Z)=0.1283_{-0.0075}^{+0.0028}$. The increasing tendency of
                fitted $\alpha_s(M_Z)$ value is associated with  the effects of  not totally  controlled within asymptotic perturbation theory expansions
                kinematical $\pi^2$ contributions to R-ratio coefficients   due to analytical continuation from the space-like to time-like energy regions.
                The applications  of the fixed orders of perturbation
                theory  expansions and  careful treatment of the
                analytical continuation effects  are commented.

\end{abstract}

%\vspace*{12.2cm}

\newpage

\section{Introduction}
\hspace*{\parindent}
The $e^+e^-$ annihilation into hadrons remains among the most informative processes of high energy physics . Its studies were  started in 1960s \cite{Gao:2021zhd} and are    continued  at present at the  experimental  facilities of the  Novosibirsk, Beijing and  Tsukuba  research  centers, even after  extensive investigations at  CERN, SLAC, and  DESY, and by  other later on $e^+e^-$ colliding machines.
The basic theoretical characteristic related to the  total cross-section of this process is
\begin{equation}
        R(s)=\frac{\sigma_{tot}(e^+e^-\rightarrow \gamma^*\rightarrow hadrons)(s)}{\sigma_0(e^+e^-\rightarrow \mu^+\mu^-)}~.
\end{equation}
where
$s=(E_{e^+}+E_{e^-})^2$ is the 
 Minkowski  squared  momentum
 transfer of the virtual photon $\gamma^{*}$  , which is equal to the square of the sum of  energies of the incoming real  electrons and positrons. In this definition
 $\sigma_0(e^+e^-\rightarrow \mu^+\mu^-)=4\pi\alpha^2/(3s)$ is the normalization factor defined  by  the leading order quantum electrodynamics (QED) massless  contribution
  to the total cross-section of  the $e^+e^-\rightarrow \gamma^*\rightarrow \mu^+\mu^-$ sub-process  and $\alpha=e^2/(4\pi)$ is the
  tree ( namely, without any radiative corrections ) expression of the QED   coupling constant, which is fixed  by the value of
  the  QED fine structure constant .
  The normalization factor is chosen in this way  to take into account  only theoretical  information
    on  only the interactions between the hadrons, created in the final states,  and between  their constituent (anti)quarks.
   While considering  R(s) ratio  it is accepted by the
 experimentalists to subtract the QED  vacuum polarization
 corrections to the propagator of real photon $\gamma$  from the total cross-section of the process  $e^+e^-\rightarrow \gamma \rightarrow{\it hadrons}$
 and divide it by the radiative QED contributions  due to photon radiations from the  initial $e^+e^-$  state  pairs.

Up to the high energy scale of the manifestation in  $e^+e^-$ annihilation of the sub-process of    creating a   virtual or real $Z$ boson, the most important contributions to   the  R ratio  come from the strong interactions of hadrons between their constituent    quarks and anti-quarks  via the gluon  exchanges.
 Different aspects related to the  determination and theoretical analysis  of  these important  effects  were discussed in  a number of recent reviews on the subject ( see, e.g.,  Refs. \cite{Logashenko:2018pcl,Jegerlehner:2003rx}).

 In general, the study of    R ratio  is one of  the most  fundamental measurements in physics of strong   interactions.   The measurements  performed at different colliders    in a wide region of energies starting from the scales of manifestation of  light
 $\rho(770)$, $\omega(782)$ and $\phi(1020)$ mesons   and above   the thresholds of production  of   heavier   J/$\Psi$ and $\Upsilon$ mesons
   confirmed   that   the constituents of  observed hadrons, namely up  $u$-,  down $d$-,  strange $s$-,  charmed $c$- and beautiful   $b$-quarks,  indeed have
 fractional electric charges and that the  special  quantum number  of quarks, i.e.   color,  is equal to   three.

In the modern  theory of strong interactions,  quantum chromodynamics or    QCD ,  it is presumed that  in addition to    quarks and
anti-quarks, there are also  the   eights of massless      gluons. Within the method of  perturbation theory,  their couplings with quarks and anti-quarks
allow one  to express    the prediction for the  R-ratio as a   power  series in  the  energy-depending QCD  coupling constant
$\alpha_s$ with  definite coefficients.  This  constant "measures" the force of  interactions  of the
 quarks and anti-quarks, bounded inside hadrons,   with  "gluing" them    gluons and obeys the property of asymptotic freedom, responsible on the inverse logarithmic
 decrease of this constant with the  increase  of energy {\it \footnote{Note, that the effects of the photon radiations from the final quark- untiquark pairs will be neglected in our considerations.}}.

In order to test the validity of   QCD,
it is necessary to extract  from different processes    the value of this coupling constant and compare the results of these extractions    between themselves
at the definite fixed reference scale. Due to the agreement between theoreticians and experimentalists,   this reference scale is fixed to be equal to the mass of the
heavy electroweak $Z$ boson $M_{Z}$,  which was
directly observed at the energies of previously working   $e^+e^-$  LEP and SLC  colliders, already  closed   Tevatron machine  and  of the continuing to collect data
Large Hadron Collider.  While fixing this scale   as a reference $\alpha_s$  scale,  the aim of   minimization of theoretical and experimental uncertainties
was   kept in mind.

In addition to  the biennial  Particle Data Group Editions (see e.g. the most recent one of Ref. \cite{ParticleDataGroup:2024cfk}),   there are a number of  detailed reviews devoted to the analysis of
 extractions of $\alpha_s$  from previous, existing and planned high energy physics experiments \cite{Deur:2016tte,
Salam:2017qdl,Pich:2020gzz,dEnterria:2022hzv} . However , in these detailed considerations  there are no discussions of the status of the extractions of $\alpha_s$ from
the R ratio experimental data. In what follows, we will fill in this gap and will concentrate on the definite comments of this topic.

The first theoretically based detailed QCD  study   was performed  in  Ref. \cite{DeRujula:1976edq}, where it
was demonstrated that   taking into account  the   positive leading order (LO)     $\alpha_s$  correcting effects,  determined independently in
    Refs. \cite{Zee:1973sr}, \cite{Appelquist:1973uz},   are really important  for
   demonstration of the realization  of the QCD asymptotic freedom property  in   the analysis of the concrete experimental data obtained  at  SLAC based SPEAR $e^+e^-$-collider in the energy region between 2 to 5 GeV and from 5 to 7.8 GeV.

This LO QCD correction  was also used in the QCD sum rules analysis of the data  of Novosibirsk VEPP-2, Orsay ADONE and Frascati DCI machines  below the 2 GeV energy region , carried out in  Ref.\cite{Eidelman:1978xy}. It
 turned out to be important for finding the correlation between the values of $\alpha_s$ and the non-perturbative  parameters, describing the effects of  condensation of  colored  gluon and quark fields in the QCD vacuum
 at rather   low energies.

A    substantiated study  of the problems
considered above problems  became possible after  the appearance
 of  specified to the theoretical   $\rm{\overline{MS}}$ -scheme expressions for the  next-to-leading order (NLO) $\alpha_s^2$ -contributions to the R ratio, calculated
analytically independently   in  Ref. \cite{Chetyrkin:1979bj} and   almost simultaneously  numerically in Ref. \cite{Dine:1979qh},
 and confirmed somewhat
 later by another
analytical  calculation of Ref. \cite{Celmaster:1979xr} .
More convincing arguments in favor of the  validity of QCD as a  theory of strong interactions were
made in the subsequent work of  Ref.\cite{Barnett:1980sm} , where these  positive NLO $\alpha_s^2$ effects were taken into account in the process of
analyzing   SLAC SPEAR   and DESY PETRA  experimental data. Moreover, after measuring the  R ratio
in the energy region between 14 to 46.8 GeV  at the  PETRA collider, the  members of
the  CELLO Collaboration
\cite{CELLO:1986ijz} obtained
 the $\rm{\overline{MS}}$- scheme related   value for$\alpha_s$, namely  $\alpha_s(34~GeV)=0.169 \pm 0.025$
 In  this way,
 the
 necessity of using information on not huge but   positive NLO  $\alpha_s^2$ QCD  contributions
 to the  R ratio  \cite{Chetyrkin:1979bj,Dine:1979qh,Celmaster:1979xr}
 for  describing concrete  experimental data within the framework of perturbative  QCD  was clarified.

 The results  obtained by the  CELLO collaboration in  Ref.\cite{CELLO:1986ijz}   were  further  confirmed in the work  \cite{Marshall:1988ri} (with its  detailed discussion  in Ref. \cite{Marshall:1989yi})
 and in the study of Ref.  \cite{DAgostini:1989tvq}, aimed at more careful $\alpha_s$-extraction after supplementing the data of the  PETRA machine with the   existing  data from  the ADONE,SPEAR, DORIS, CESR, PEP and TRISTAN  $e^+e^-$ colliders. Another aim of these works was the inclusion into the studies of   theoretical information on the
  next-to-next-to-leading order (NNLO) $\alpha_s^3$ QCD effects, obtained from the work \cite{Gorishnii:1988bc} with the unfortunate dramatic results.
 Among them was the finding  that  the  result of  Ref.\cite{Gorishnii:1988bc} for the $\alpha_s^3$ coefficient  used in these studies  turned out to be   erroneous.

 The {\bf  corrected  negative  NNLO $\alpha_s^3$ } analytical  contribution to the  R ratio was obtained  in  Ref. \cite{Gorishnii:1990vf} after more than a   two-year  long
 project. It   agreed  with the related one,  given in  Ref.\cite{Surguladze:1990tg}. Later on,  it was also confirmed by
 an    independent calculation of Ref. \cite{Chetyrkin:1996ez}. The  re-evaluation project was indeed rather important in view of approaching
 era of high energy SLAC SLC and CERN LEP colliders. Almost immediately after   publication,  the results  of
 Refs.\cite{Gorishnii:1990vf,Surguladze:1990tg} were implemented in  Ref.\cite{Branchina:1992vy}  to the analyze  $e^+e^-\rightarrow {hadrons}$
data  from   PETRA, PEP, TRISTAN and other colliders
  (including  LEP). As a  result
  a high value $\alpha_s(M_Z) = 0.148\pm 0.019~(exp) \pm 0.005~(theor)$  was obtained,  which was  surprising  even to the authors of this work
  \footnote{To avoid the discussions of the effects of electro-weak corrections we will not
 consider the further on  determinations  of   $\alpha_s(M_Z)$   from LEP data }.

 In the work of CLEO collaboration from CESR   the R measurement at $\sqrt{s}=10.52$ GeV energy scale
 was performed \cite{CLEO:1997eca}. This implies  the following   NNLO  value of the QCD coupling constant
 $\alpha_s(M_Z)=  0.13 \pm 0.005 ~(stat) \pm 0.03~(syst)$ \cite{CLEO:1997eca}. The performed later more detailed analysis of this data with taking into account
 charm quark mass effects gave the following more detailed result $\alpha_s(M_Z)=0.124^{+0.013}_{-0.014}$\cite{Kuhn:2001dm}.

 The further on measurements , performed with the help of CLEOIII detector allowed to get 7 more points for R in the energies between 6.964 GeV and 10.538 GeV
 and the following QCD coupling constant NNLO  value $\alpha_s(M_Z)=0.126 \pm 0.005^{+0.015}_{-0.011}$ \cite{CLEO:2007suf}.  However, the authors of \cite{Kuhn:2007tc}
 claimed that the inclusion of charm quark  mass effects and of the proper matching between the effective theories with four and five flavours changes
 this result to $\alpha_s(M_Z)=0.110^{+0.010+0.010}_{-0.012-0.011}$ later on \cite{Kuhn:2007tc}.
Another     interesting,  though unpublished,  combined  NNLO QCD   analysis
        of the  $e^+e^-$ data   without the CLEOIII data  of Ref.\cite{CLEO:2007suf}
resulted in a  relatively larger value of $\alpha_s(M_Z)$ , namely    $\alpha_s(M_Z)=0.128 \pm 0.032$ \cite{Zenin:2001yx}.

In the process of determination of the  world  average  combined value  $\alpha_s(M_Z)=0.1180\pm 0.0009$
Particle Data Group      \cite{ParticleDataGroup:2024cfk} used  another   $e^+e^-$ R- ratio related
NNLO   fixed order perturbation theory result $\alpha_s(M_Z)=0.1158\pm 0.0022$, obtained in
Ref. \cite{Boito:2018yvl} from the QCD finite energy sum rules analysis of the R-ratio low energy data to 2 GeV. It was also
commented there that the data above 2 GeV did not provide much additional constraint on this analysis.

Not long ago,   new more detailed  experimental information on R-ratio values  in this  low energy region
 $1.8~GeV \leq \sqrt{s} \leq 3.7$ GeV  was provided by KEDR collaboration from the   Novosibirsk   VEPP-4M collider   \cite{Anashin:2015woa, Anashin:2016hmv,KEDR:2018hhr} and   by BESIII collaboration from the  Beijing   BEPC-II collider
  \cite{BESIII:2021wib} in the region 2.23 GeV to 3.67 GeV.  These data were used recently
  in the work \cite{Boito:2025qfz} in the process of
  taking into account rather sizable  negative next-to-next-to-next-to-leading order ( N$^3$LO)
  $\alpha_s^4$ contributions to the R ratio, analytically obtained in \cite{Baikov:2008jh} and confirmed in \cite{Herzog:2017dtz}.

 In the study  performed in \cite{Boito:2025qfz},  the
 problems of taking into account  not yet calculated  next-to-next-to-next-to-next-to-leading   order (N$^4$LO) $\alpha_s^5$ effects were also considered.
 First estimates of these terms were obtained   in \cite{Baikov:2008jh}
 by a theoretical procedure,
  developed in Ref.\cite{Kataev:1995vh} . In Ref.\cite{Boito:2025qfz} the results of further  more detailed Pad\'e
 related estimates of Ref.\cite{Boito:2018rwt}  were applied , which within specified theory uncertainties
 turned out to be in  agreement with the number of of Ref.\cite{Baikov:2008jh}
 and were also recently confirmed in Ref.\cite{Abbas:2026zma}  by another  approach.

  In  Ref.\cite{Boito:2025qfz}  the $\alpha_s$ value was {\it not extracted} but {\it fixed} from the  world
  average value  $\alpha_s(M_Z)=0.1180\pm 0.0009$, given by the
  Particle Data Group      \cite{ParticleDataGroup:2024cfk}. In  our
  work,  we will follow the way back and  consider the problem of extraction of $\alpha_s$  from  the analysis of  available now  R ratio data  of the  KEDR collaboration  \cite{Anashin:2015woa, Anashin:2016hmv,KEDR:2018hhr}
  and BESIII  collaboration  \cite{BESIII:2021wib} in the energy
  region below the thresholds of production  of charm quark- and  anti-quark pairs and
  {\bf  truncating} R-ratio  QCD   expression {\bf at each subsequent
  order} of perturbation theory expansion.

 While performing  this study, we will try to achieve the following aims :
\begin{itemize}
\item  to get    new $e^+e^-$-related   $\alpha_s(M_Z)$ values for the fundamental constant of  strong interactions from the
  independent analysis of these   recent  experimental data for the  R ratio   ;
 \item to take into account maximum information on the   theoretical perturbative  QCD approximations   for both
 experimentally  measured   R ratio  and for    the  energy evolution  of strong interactions   coupling constant $\alpha_s$,
  obeying the QCD asymptotic freedom property;
 \item to clarify whether the analysis of  existing new lower  energy experimental $e^+e^-$ annihilation  data
  is sensitive to  the  {\bf  difference between the sign
 structures  and the values of the  QCD  perturbation theory coefficients}, originally defined  in the Euclidean region,  and the ones  further on   transformed into
the  physical  Minkowski region, where the experimental  data for the R ratio are  obtained.
\end{itemize}

We will also try to get personal impression on the conclusion made in Ref.\cite{Boito:2025qfz}  that the QCD fits give
a definite  preference to  KEDR data in relation to  BESIII ones for the $e^+e^-$ to hadrons total cross sections and
whether it is possible to minimize the observed in Ref.\cite{Boito:2025qfz} definite tension between perturbative QCD predictions for R-ratio
and the BESIII data by applying definite kinematical   cuts to these data.

 These cuts disfavor the inclusion into final physical fits  of the extracted 
BESIII 8  data points for  R above 3.4 GeV. We will see , that 
 taking them into the account  result in the significant decrease of 
the goodness-of-fit probability of the outcomes of the fits   significantly  below the commonly accepted   $5\%$ level. 

 Consequences of  underestimating   theoretical uncertainties  of the
   N$^3$LO QCD extraction of $\alpha_s$ from  KEDR data  \cite{Shen:2023qgz}, where these uncertainties   were fixed by choosing  the proposed in Ref.\cite{Brodsky:2013vpa}, but   constructively criticized  in Refs.\cite{Kataev:2013vua,Kataev:2014jba,Kataev:2023xqr,
        Mikhailov:2024mrs}  way of fixing them  will also  be commented.

\section{Theory related setup }

In theory the expression for  the R ratio defined in the  Minkowski time-like region   is determined by the imaginary part of the photon vacuum polarization function transformed to the complex region

\begin{equation}
        R(s)=12\pi Im \Pi (-s+i\epsilon)
\end{equation}
which enters into  the expression of the defined in the Euclidean region two-point Green function of the related quark currents
\begin{equation}
        \Pi_{\mu\nu}(q)=i \int e ^{iqx}<0|T[J_{\mu}^{em}(x) J_{\nu}^{em}(0)]|0>d^4x=(-g_{\mu\nu}q^2+q_{\mu}q_{\nu})\Pi(Q^2)
\end{equation}
where the Euclidean momentum transfer is $Q^2=-q^2$ and  $J_{\mu}^{em}(x)=\sum Q_q\overline{ \psi}_q(x)\gamma_{\mu}\psi_q(x)$ is the vector current of quarks with $Q_q$ fractional charges in units of electron charge. Taking the derivative of the photon vacuum polarization function $\Pi(Q^2)$, it is possible to get the following relations
between the  characteristics of the $e^+e^-$ annihilation to hadrons in the Euclidean and Minkowski regions :

\begin{equation}
        \label{relations}
D(Q^2)= -12\pi^2 Q^2\frac{d\Pi(Q^2)}{dQ^2}=Q^2 \int_0^{\infty} \frac{R(s)}{(s+Q^2)^2}ds ~~~,~~~ R(s)=\frac{1}{2\pi i} \int_{-s-i\epsilon}^{-s+i\epsilon} dQ^2 \frac{D(Q^2)}{Q^2}~~.
\end{equation}

In the case of three flavours of light active quarks and within the  perturbative QCD,  the introduced above quantities can be expressed as
\begin{equation}
        \label{R}
        R_{uds}(s)= 3\sum_{q=u,d,s} Q_q^2 \big[1+r_1a_s(s)+
        \sum_{k\geq 2}r_k(f)a_s(s)^k\big]
\end{equation}
\begin{equation}
        \label{D}
D(Q^2)=3\sum_{q=u,d,s}Q_q^2 \big[1+ d_1a_s(Q^2)+\sum_{k\geq 2}d_k(f)a_s(Q^2)^k\big]
\end{equation}
where   $f$ is a number of  quarks flavours, $a_s(s)=\alpha_s(s)/\pi$ and  $a_s(Q^2)=\alpha_s(Q^2)/\pi$ are  the expansion parameters with the related running quark-gluon  coupling constants,  defined in the
Minkowski and Euclidean regions respectively.  Since in the light-quarks sector $(\sum_{q=u,d,s} Q_q)^2=(\frac{2}{3}-\frac{1}{3}-\frac{1}{3})^2=0$, the
proportional to this structure extra contributions of  order $O(a_s^3)$
are not considered. In the Euclidean region, the energy-dependence of the QCD coupling constant is defined by the solution of
the  following  renormalization-group equation :
\begin{equation}
        Q^2\frac{\partial{a_s(Q^2)}}{\partial Q^2}=\beta(a_s)=-\sum_{n\geq 0}\beta_na_s^{n+2}~~~.
\end{equation}
In the  $\rm{\overline{MS}}$-scheme, usually considered as the reference scheme,  in the
 energy region below the production of charm-anti-charm pairs,  the related expressions for the QCD coupling constant, expressed through the  order-by-order  expansion
 in the inverse logarithmic terms,  up to the N$^4$LO  we will be interested in,  can be written down as

\begin{eqnarray}
        \label{NLO}
      a_s^{NLO}(Q^2)&=&\frac{1}{\beta_0 L_{Q^2}}-\frac{\beta_1ln L_{Q^2}}{\beta_0^3L_{Q^2}^2} ~~~\\ \label{NNLO}
           a_s^{NNLO}(Q^2)&=&a_s^{NLO}(Q^2)+\Delta_{NNLO}~~~\\ \label{N3LO}
    a_s^{N^3LO}(Q^2)&=&a_s^{NNLO}(Q^2)+\Delta_{N^3LO} \\ \label{N4LO}
            a_s^{N^4LO}(Q^2)&=&a_s^{N^3LO}(Q^2)+\Delta_{N^4LO}~~~.
\end{eqnarray}
The
related high-order terms in these  expressions   are defined as
\begin{equation}
        \Delta_{NNLO}=\frac{\beta_1^2(l_{Q^2}^2-l_{Q^2}-1)+\beta_0\beta_2}{\beta_0^5L_{Q^2}^3}~~~,
\label{NNLO}
\end{equation}
\begin{equation}
        \Delta_{N^3LO}=\frac{\beta_1^3(-2l_{Q^2}^3+5l_{Q^2}^2+4l_{Q^2}-1)-6\beta_0\beta_2\beta_1l_{Q^2}+\beta_0\beta_3}{2\beta_0^7L_{Q^2}^4}~~~,
\label{N^3LO}
\end{equation}
\begin{eqnarray}
        \Delta_{N^4LO}&=&\frac{18\beta_0\beta_2\beta_1^3(2l_{Q^2}^2-l_{Q^2}-1)+\beta_1^4(6l_{Q^2}^4-26l_{Q^2}^3-9l_{Q^2}^2+24l_{Q^2}+7)}{6\beta_0^9L^5_{Q^2}} \\ \nonumber
        &-& \frac{\beta_0^3\beta_1\beta_3(12l_{Q^2}+1)-10\beta_0^3\beta_2^2-2\beta_0^4\beta_4}{\beta_0^9L_{Q^2}^5}
\label{N^4LO}
\end{eqnarray}
where $L_{Q^2}=ln(Q^2/\Lambda_{\rm{\overline{MS}}}^{(f=3)2})$, $l_{Q^2}=lnL_{Q^2}$, 
$\Lambda_{\rm{\overline{MS}}}^{(f=3)}$ is the corresponding scale parameter of the world with $f=3$ number of active quarks flavours, subsequently extracted at each   order of perturbation theory .  The  numerical expressions  for  the considered coefficients of the QCD $\beta$-function  read
\begin{eqnarray}
        \beta_0&=& 2.75-0.16667f ~~;~~\beta_1=6.375-0.79167f~~  \\ \nonumber
        \beta_2&=&22.320-4.3689f+0.094039f^2 \\ \nonumber
        \beta_3&=&114.23-27.134f+1.5824f^2+0.0058567f^3 \\ \nonumber
        \beta_4&=&524.558-181.80f+17.156f^2-0.22586f^3-0.0017993f^4~~.
\end{eqnarray}
In the analytical form,  the  first renormalization-scheme independent  LO coefficient $\beta_0$ was  obtained  at  the same time in the works  \cite{Gross:1973id,Politzer:1973fx} . Its discovered  positive
value served as a
the cornerstone for understanding that QCD obeys the property of asymptotic freedom. Within gauge-invariant schemes, the second NLO
coefficient $\beta_1$  is also scheme-independent and was also simultaneously calculated in two works \cite{Jones:1974mm,Caswell:1974gg}. It was  confirmed  later on in
Ref.\cite{Egorian:1978zx} during    analytical calculations of the in generally  scheme-dependent $\beta_2$ coefficient
completed in Ref.\cite{Tarasov:1980au}. The result of Ref.\cite{Tarasov:1980au}
 was confirmed later on in  Ref.\cite{Larin:1993tp}.  It is important for getting self-consistent
information on the energy-dependence of the NNLO approximation of $\alpha_s$ . The next N$^3$LO term was analytically evaluated  in
 Ref.\cite{vanRitbergen:1997va} and confirmed seven years later in Ref.\cite{Czakon:2004bu}.  The fifth $\beta_4$-coefficient was calculated in
Ref.\cite{Baikov:2016tgj } and was confirmed in Refs.\cite{Herzog:2017ohr,Luthe:2017ttg}.  In our $e^+e^-$ to hadrons related
consideration, it will become important in the process of taking into account of  the estimated
 N$^3$LO  $\alpha_s^5$-contribution to the  R ratio.

We will consider the QCD coupling constant  evolution  in the  time-like domain after the redefinition of  $\alpha_s(Q^2)$ to
 the Minkowski time-like  region and redefining perturbation theory expansion in powers of $\alpha_s(s)$ to the expansion in powers of
 $1/L_s=1/ln(s/\Lambda_{\overline{MS}}^{(f=3)2})$ -function.
 This redefinition will be made by  changing  $1/L_Q$ to
$1/L_s$ in Eqs.(\ref{NLO})-(\ref{N4LO}) above.  In general, this shift  leads to the appearance of additional $\pi^2$-terms
in the time-like
domain. They
 arise  from the theoretical expression of the  time-like  version of the  QCD coupling constant
 as the result of using  the
following analytical continuation procedure : $1/L_Q^{2n}\rightarrow1/( L_s \pm i\pi)^{2n}$. These effects were considered  previously in
Refs. \cite{Krasnikov:1982fx,Radyushkin:1982kg,Pennington:1983rz} and were analyzed within the context of QED in the most recent work of Ref.\cite{Krasnikov:2026cxw}.

In our further on studies  we will  define the QCD expansion parameter in the Minkowski region as it was defined by Eqs.(8)-(14) in the Euclidean region  and  trace
extra proportional to powers of
$\pi^2$ analytical continuation    contributions
  in the  coefficients for the  R ratio
\begin{equation}
r_k(f)=d_k(f)-\Delta_k(f)~~~.
\end{equation}
These analytical continuation effects have
the following form :
\begin{eqnarray}
        \Delta_1&=&\Delta_2=0 ~~,~~~
        \Delta_3(f) = \frac{\pi^2\beta_0^2}{3}d_1 \\ \nonumber
        %=24.9-3.02f-0.091f^2 \\ \nonumber
        \Delta_4(f)&=& \pi^2\big[\beta_0^2 d_2(f)+\frac{5}{6}\beta_0\beta_1d_1\big]
        %=292.4-53.2f+2.67f^2-0.032f^3
        \\ \nonumber
        \Delta_5(f)&=&\pi^2\big[2\beta_0^2d_3(f)+\frac{7}{3}\beta_0\beta_1d_2(f)+\frac{1}{2}\beta_1^2d_1+\beta_0\beta_2d_1\big]-\frac{\pi^4}{5}\beta_0^4d_1
\end{eqnarray}

 As was shown in an   elegant way  in Ref.\cite{Bjorken:1989xw}, in   general these analytical continuation effects arise  after     taking the following integral  even powers of the
the   renormalization-group
controllable logarithmic terms:
\begin{equation}
        \int_0^{\infty}dx\frac{ln^{2n}(x)}{(x+1)^2}= 2(1-2^{1-2n})\zeta_{2n} (2n)!  ~  n \geq 1~~.
\end{equation}
where $\zeta_{2n}=\sum_{k\geq 1}^{\infty}(1/k)^{2n}$ is the Riemann zeta function with its even values fixed as $\zeta_2=\pi^2/6$, $\zeta_4=\pi^4/90$, etc.

Their influence on   the  change of the  asymptotic structure  of the perturbative expansion series for the Euclidean Adler   $D$-function of Eq.(\ref{relations})
 after analytical continuation to
the Minkowski region was considered   in Ref.\cite{Kataev:1995vh} and  later in
Refs.\cite{Bakulev:2010gm,Nesterenko:2017wpb,AlamKhan:2023dms}. From a  mathematical point of view,  the effects of these terms were studied in Ref.\cite{Gracia:2023qdy}. They can be summed up using either the
 contour-improved perturbation theory (CIPT) \cite{Pivovarov:1991rh,LeDiberder:1992jjr},    or   theoretically equivalent to it
analytical perturbation theory (APT) developed in   Refs.\cite{Shirkov:1997wi,Shirkov:2006gv} .

 In what follows, we will use the language of the fixed order perturbation theory, which presumes the existence of
the unique expansion parameter $\alpha_s$ in the following
 $\rm{\overline{MS}}$-scheme perturbation theory series in the  Euclidean and Minkowski regions:
\begin{equation}
        \label{D}
        D(Q^2)=3\sum_{q=u,d,s} Q_{q}^2 \big[1+ a_s+1.6398 a_s^2+6.3710a_s^3+49.076a_s^4+\sim 275a_s^5+O(a_s^6)\big]
\end{equation}
\begin{equation}
        \label{R}
        R(s)= 3\sum_{q=u,d,s} Q_q^2 \big[1+a_s+1.6398 a_s^2-10.2839 a_s^3-106.8798 a_s^4- \sim 505 a_s^5+O(a_s^6)\big]~~.
\end{equation}
Notice the change of signs in the corresponding coefficients starting from the $O(a_s^3)$ level up to at least $O(a_s^5)$ terms after taking into account the  effects
of analytical continuation.  The corresponding truncated massless expressions will be used in the process of fits of the concrete experimental data
with the aims of defining first the values of the parameter $\Lambda_{\rm{\overline{MS}}}^{(f=3)}$ and then the related   $\alpha_s(m_{\tau})$  values.

To get  the related  $\alpha_s(M_Z)$ values,   we   first determine  the  values of $\Lambda_{\rm{\overline{MS}}}^{(f=5)}$  by transforming  the
 results for   $\Lambda_{\rm{\overline{MS}}}^{(f=3)}$ , obtained from the fits,
        to the world with $f=5$ numbers of active flavours.
        This will be carried out   by passing the  thresholds of production in the $e^+e^-$  annihilation into hadrons process  of thresholds of the  productions of pairs of   $c$ and $b$ quarks,
        fixed by the  masses of the  related mesons   $M_{D^0}=1.86$ GeV and $M_{B}=5.28$ GeV.
        This will be done   successively  in  every order of perturbation theory , based on the results of the analysis  carried out 
        in \cite{Kataev:2001kk}, in accordance with the formulas given below  
                \begin{equation}
        \beta_0^{(f+1)}
        ln\frac{\Lambda^{(f+1)2}_{\overline{MS}}}{\Lambda^{(f)2}_{\overline{MS}}}
        =\beta_0^{(f)}L_h\bigg[ \bigg(\frac{\beta_0^{(f+1)}}{\beta_0^{(f)}}-1\bigg)
        +\delta_{NLO}+\delta_{N^2LO}+\delta_{N^3LO}
        +\delta_{N^4LO}\bigg]
\end{equation}
where $L_h=ln(M_h^2/\Lambda^{(f)2}_{\overline{MS}})$ depends on  the matching scales $M_h=M_{D_0}$ and$ M_{h}=M_{B}$ when $f=3$ or $f=4$ and
the high order perturbative QCD contributions are fixed as
\vspace{0.5cm}
\begin{equation}
        \delta_{NLO}=\bigg(\frac{1}{\beta_0^{(f)}L_h}\bigg)\bigg[\bigg(\frac{\beta_1^{(f+1)}}{\beta_0^{(f+1)}}-\frac{\beta_1^{(f)}}{\beta_0^{(f)}}\bigg)lnL_h-\frac{\beta_1^{(f+1)}}{\beta_0^{(f+1)}}ln\frac{\beta_0^{(f+1)}}{\beta_0^{(f)}}\bigg]
        \end{equation}
        \begin{equation}
                \delta_{NNLO}=\bigg(\frac{1}{\beta_0^{(f)}L_h}\bigg)^2\bigg[\frac{\beta_1^{(f)}}{\beta_0^{(f)}}\bigg(\frac{\beta_1^{(f+1)}}{\beta_0^{(f+1)}}-
                \frac{\beta_1^{(f)}}{\beta_0^{(f)}}\bigg)lnL_h+\bigg(\frac{\beta_1^{(f+1)}}{\beta_0^{(f+1)}}\bigg)^2-\bigg(\frac{\beta_1^{(f)}}{\beta_0^{(f)}}\bigg)^2
                -\frac{\beta_2^{(f+1)}}{\beta_0^{(f+1)}}+\frac{\beta_2^{(f)}}{\beta_0^{(f)}}-C_2\bigg]
        \end{equation}
        \begin{eqnarray}
                \delta_{N^3LO}&=&\bigg(\frac{1}{\beta_0^{(f)}L_h}\bigg)^3\bigg[ -\frac{1}{2}\bigg(\frac{\beta_1^{(f)}}{\beta_0^{(f)}}\bigg)^2\bigg(\frac{\beta_1^{(f+1)}}{\beta_0^{(f+1)}}-
                \frac{\beta_1^{(f)}}{\beta_0^{(f)}}\bigg)ln^2L_h \\  \nonumber
                &+&\frac{\beta_1^{(f)}}{\beta_0^{(f)}}\bigg[-\frac{\beta_1^{(f+1)}}{\beta_0^{(f+1)}}\bigg(\frac{\beta_1^{(f+1)}}{\beta_0^{(f+1)}}-
                \frac{\beta_1^{(f)}}{\beta_0^{(f)}}\bigg)+\frac{\beta_2^{(f+1)}}{\beta_0^{(f+1)}}-\frac{\beta_2^{(f)}}{\beta_0^{(f)}}+C_2\bigg]lnL_h \\ \nonumber
                &+&\frac{1}{2}\bigg(-\bigg(\frac{\beta_1^{(f+1)}}{\beta_0^{(f+1)}}\bigg)^3-\bigg(\frac{\beta_1^{(f)}}{\beta_0^{(f)}}\bigg)^3
                -\frac{\beta_3^{(f+1)}}{\beta_0^{(f+1))}}+\frac{\beta_3^{(f)}}{\beta_0^{(f)}}\bigg) \\ \nonumber
                &+&\frac{\beta_1^{(f+1)}}{\beta_0^{(f+1)}}\bigg(\bigg(\frac{\beta_1^{(f)}}{\beta_0^{(f)}}\bigg)^2+\frac{\beta_2^{(f+1)}}{\beta_0^{(f+1)}}-
                        \frac{\beta_2^{(f)}}{\beta_0^{(f)}}+C_2\bigg)-C_3\bigg]~~.
\end{eqnarray}
The expressions for $\delta_{NLO}$, $\delta_{NNLO}$ and $\delta_{N^3LO}$ follow from the  related calculations of Refs.\cite{Bernreuther:1981sg,Larin:1994va,Chetyrkin:1997un,Chetyrkin:2005ia}.
At  N$^4$LO, the corresponding correction was  also explicitly evaluated in Ref.\cite{Kniehl:2006bg}
$\delta_{N^4LO}=\delta(\{\beta_i^{(f)},\beta_i^{(f+1)}\},C_2,C_3,C_4)/(\beta_0^{(f)}L_h)^4$ with $\{\beta_i^{(f)},\beta_i^{(f+1)}\}=(\beta_0^{(f)},\beta_0^{(f+1)},\dots,\beta_i^{(f)},\beta_i^{(f+1)})$ and $1\leq i\leq 4$ and $C_2,C_3,C_4$ are the concrete coefficients evaluated in Refs.\cite{Larin:1994va,Chetyrkin:1997un} and Ref.\cite{Kniehl:2006bg} respectively.
The  application of this expression  in  concrete  considerations did not give   serious  N$^3$LO changes in the   $\Lambda_{\rm{\overline{MS}}}^{(f=5)}$ values.
In view of this and to save the space,  we will not present the result of  Ref.\cite{Kniehl:2006bg} in our work explicitly.

It should be noted that due to minor  differences between
        the realizations of the perturbation theory expansions in
        terms of powers of used by us  $1/L_Q$ or $1/L_s$ expansions  and more oftently
        considered  $\alpha_s$  expansions   and of the  realizations of the related  matching conditions
        there are some differences     between the final results of
        applications of the related matching procedures
based on the formulas mentioned above and those obtained within the
framework of the RunDec package of Refs.
\cite{Chetyrkin:2000yt,Herren:2017osy}. However, these  differences
 do not exceed $2\times 10^{-4}$ in $\alpha_s(M_Z)$-values  and we will  neglected them  in our further  considerations.

\section{Experimentally related setup.}

Let us consider first the data of the   KEDR Collaboration  data for
the  R ratio in the region from 1.84 GeV to 3.72 GeV, given in
Refs.\cite{Anashin:2015woa,  Anashin:2016hmv, KEDR:2018hhr}.
In our work we use the $R_{uds}$  values published in \cite{KEDR:2018hhr},
which contains measurement results for 22 energy points, including the
corresponding statistical and systematic uncertainties,
as well as correlation matrix for systematic uncertainties.
This information allows one to determine the covariance matrix for KEDR
data as
\begin{equation}
 C_{ij}^{KEDR}=\delta_{ij}\sigma_{stat,i}^2+\sigma_{syst,i} \rho_{ij} \sigma_{syst,j}
\label{eq:KEDRerr}
\end{equation}
where  $\delta_{ij}$ is the  Kronecker delta. Here  $\sigma_{stat,i}$ and $\sigma_{syst,j}$
are statistical  and systematic errors and  $\rho_{ij}$ is the corresponding  correlation matrix, explicitly given in \cite{KEDR:2018hhr}.

Let us now   define the  corresponding $\chi_0^2$ minimization function
as

\begin{equation}
        \chi^2_{0}= \sum_{i}\sum_{j} \bigg(R^{{exp}}(s_i)-R^{{th}}(s_i)\bigg) (C^{-1})_{ij} \bigg(R^{{exp}}(s_j)-R^{{th}}(s_j)\bigg)~,
        \label{Chi2:1}
\end{equation}

where $R^{{exp}}(s_i)$ and  $R^{{th}}(s_i)$ are the measured and theoretically defined values of the  R  ratio , fixed at the concrete
energy points,  and
$C^{-1}_{ij}$ are the  coefficients of the inverse covariance matrix.

In  the publication of the  BESIII Collaboration \cite{BESIII:2021wib},  their data on
 14 experimental points for the  R ratio in the related  energy region
 was given but the information on the corresponding
 correlation matrix was not explicitly specified.
 In our analysis,  we  construct this correlation matrix
 $C_{ij}^{BESIII}$  using the experimental data,
 given  in "Table I" of this work .

Taking into account the
 comments on  systematic uncertainties of BESIII data points for R ,given  in the related "Table I" caption, namely
 the information  that
  "uncertainties from the last four sources" (from luminosity, trigger efficiency, signal model and initial state radiation) "are
 correlated between the energy points",  we  estimate the corresponding correlated  systematical
 uncertainty  $\sigma_{syst. corr.}$ for every presented BESIII R ratio point, summing these four errors in quadrature.
 The remaining   three uncertainties for event selection, QED background and beam background are used to fix
 the uncorrelated systematical error $\sigma_{syst. uncorr.}$. The
 results of our estimates are given in Table \ref{Tab:BESdata} below.

\renewcommand{\arraystretch}{1.1}
\begin{table}[H]
        \begin{center}

                        \caption { \label{Tab:BESdata} BESIII data points
                                used in the fit with total statistical and systematical uncertainties. The second and third columns contain estimated by us  uncorrelated and
                               correlated systematic uncertainties.
                               }
                        \begin{tabular}[l]{|l|c|c|c|} \hline
                                $\sqrt{s}$, GeV  &  R(s) &
                                $\sigma_{syst. uncorr.}$ &
                                $\sigma_{syst. corr.}$    \\\hline
                                2.2324  &$2.286 \pm 0.008 \pm 0.037$     &  0.0125  & 0.0349 \\
                                2.4000  &$2.260 \pm 0.008 \pm 0.042$     &  0.0143  & 0.0397 \\
                                2.8000  &$2.233 \pm 0.008 \pm 0.055$     &  0.0163  & 0.0531 \\
                                3.0500  &$2.252 \pm 0.004 \pm 0.052$     &  0.0181  & 0.0492 \\
                                3.0600  &$2.255 \pm 0.004 \pm 0.054$     &  0.0189  & 0.0506 \\
                                3.0800  &$2.277 \pm 0.003 \pm 0.046$     &  0.0179  & 0.0424 \\
                                3.4000  &$2.330 \pm 0.014 \pm 0.058$     &  0.0174  & 0.0554 \\
                                3.5000  &$2.327 \pm 0.010 \pm 0.062$     &  0.0217  & 0.0579 \\
                                3.5424  &$2.319 \pm 0.006 \pm 0.060$     &  0.0165  & 0.0575 \\
                                3.5538  &$2.342 \pm 0.008 \pm 0.064$     &  0.0194  & 0.0613 \\
                                3.5611  &$2.338 \pm 0.010 \pm 0.066$     &  0.0206  & 0.0623 \\
                                3.6002  &$2.339 \pm 0.006 \pm 0.065$     &  0.0194  & 0.0618 \\
                                3.6500  &$2.352 \pm 0.009 \pm 0.067$     &  0.0221  & 0.0628 \\
                                3.6710  &$2.405 \pm 0.010 \pm 0.067$     &  0.0234  & 0.0623 \\\hline
                        \end{tabular}
                \end{center}
\end{table}
Taking these considerations into account,  we now construct the corresponding covariance matrix of BESIII data :
\begin{equation}
C_{ij}^{BESIII}=\delta_{ij}(\sigma_{stat,i}^2+\sigma_{syst. uncorr.,i}^2) + \sigma_{syst. corr.,i}\sigma_{syst. corr.,j}.
\label{eq:BESmat}
\end{equation}

In general, both  BESIII or KEDR or  any other
experimental data  may deviate from the expected theoretical expressions.
In R measurements, this may be due to the  systematic errors say in the luminosity measurements or  efficiency determinations.
We will take these possible deviations into account in the following way :
\begin{itemize}
        \item   for all experimental points, the minimum correlated systematic uncertainty $\sigma_0$  is determined;
        \item  this value is quadratically subtracted from the systematic uncertainties
for each point under consideration;
\item a new covariance matrix $\tilde{C}_{ij}$ is determined ;
\item the $\chi^2$ minimization function is redefined as
\end{itemize}

\begin{equation}
\chi^2_1=\frac{(\nu-1)^2}{\nu^2\sigma_0^2} +\frac{1}{\nu^2} \sum_{i}\sum_{j} \bigg(\nu R^{{exp}}(s_i)-R^{{th}}(s_i)\bigg)(\tilde{C}^{-1})_{ij} \bigg(\nu R^{{exp}}(s_j)-R^{{th}}(s_j)\bigg),
        \label{Chi2:2}
\end{equation}

where $\nu$ is a  free
 normalization parameter, which reflects,  to a certain extent,  how
 much, on average, the experimental data differ from the theoretical dependence
 over the entire energy range under consideration. A minimization function based on  only the $\chi^2_0$ function can lead to
 biased results, since the theoretical curve cannot move over the entire
 range of measured experimental values of R.
 Using $\chi^2_1$ allows us to separate the overall bias in the R data from
 the shape of the  theoretical  curve, which is actually compared in the fit with its
 theoretical energy dependence.

 This $\chi_1^2$ redefinition procedure was applied previously in Refs.\cite{Marshall:1988ri,Marshall:1989yi} in the process of   NLO fits of R ratio
 data above the  thresholds of production of $c$ and $b$  quarks. In our work,  we  use both the  $\chi_0^2$ and $\chi_1^2$ minimization functions
 in the process of   fitting $e^+e^-$ hadrons data below  the thresholds
 of production of charm quarks.
 We consider  the approaches based on  using minimization functions that take into
 account  possible overall bias as rather useful for estimating extra  theory related experimental  uncertainties
 . Note, that the introduced parameter $\nu$ of the  function $\chi^2_1$ is directly
 related to the systematic uncertainties in more careful determination of  the possible uncertainties  related to those ones due to variations of
 efficiency and luminosity.

\section{Results of fits of KEDR data }
Let us first apply  the considerations of the previous Section for analyzing   KEDR R(s)  experimental  data  given in Ref.\cite{Anashin:2015woa,  Anashin:2016hmv,KEDR:2018hhr}. In the process of   concrete fits,
 made with the help of the standard Root  MINUIT library, the  truncated at the NLO, NNLO, N$^3$LO and N$^4$LO    perturbative QCD
 expression for the   R ratio (see Eq.(\ref{R})) and the analogously
 truncated inverse logarithmic expressions for   the QCD coupling constant $\alpha_s$
 %(see Eqs.(\ref{NLO})-(\ref{N^4LO}))
 were used.
 %The results of our  analysis are presented in Table 2 below.
 We do not present errors for results obtained in  $\alpha_s^5$ order
 since the numerical value for a given contribution to R is only an estimate.
\begin{table}[H]
        \begin{center}
                \caption{\label{Tab:Results1}
                        Results of fitting KEDR data  using two different $\chi^2$
                        minimization functions for the $O(\alpha_s^2)$ , $O(\alpha_s^3)$ NNLO, $O(\alpha_s^4)$ N$^3$LO QCD approximations of R(s)  and
                        with taking into account the estimated
                        $O(\alpha_s^5)$ N$^4$LO correction.
                        P($\chi^2$) represents the  p-value (goodness-of-fit probability) for a fit.}

                \begin{tabular}[c]{|c|c|c|c|c|c|}   \hline R(s)  approximation
                        & $O(\alpha_s^2)$ & $O(\alpha_s^3)$ & $O(\alpha_s^4)$ & $O(\alpha_s^5)$ estimate \\ \hline  \hline
                        $\chi_0^2 (KEDR)/ndf$       &
                        5.941/21 & 5.454/21
                        & 3.677/21 & 4.448/21\\
                        P($\chi^2$), $\%$& $99.947$ &$99.973$ &$99.999$ & $99.995$\\\hline \hline
                        $\Lambda_{\rm{\overline{MS}}}^{(f=3)}$, MeV&
                        $447_{-115}^{+104}$   &
                        $490_{-137}^{+134}$  &
                        $672_{-135}^{+59}$ &
                        $605_{-127}^{+67}$\\ \hline

                        $\alpha_s(m_\tau)$ &
                        $0.3588_{-0.0612}^{+0.0663}$  &
                        $0.4034_{-0.0790}^{+0.0990}$  &
                        $0.5793_{-0.1319}^{+0.0825}$  &
                        $0.485$ \\ \hline
                        $\alpha_s(M_Z)$ ($NLO)$
                        &$0.1189_{-0.0062}^{+0.0049}$   & &  &
                        \\
                        $\Lambda_{\rm{\overline{MS}}}^{(f=5)}$, MeV &
                        $238_{-72}^{+70}$ &  &  &  \\
                        \hline
                        $\alpha_s(M_Z)$ ($NNLO)$&    &
                        $0.1226_{-0.0073}^{+0.0061}$ &                 &
                        \\ $\Lambda_{\rm{\overline{MS}}}^{(f=5)}$, MeV
                        &
                        & $267_{-88}^{+94}$ &  &  \\ \hline
                        $\alpha_s(M_Z)$ ($N^3LO)$&  &
                       &     $0.1305_{-0.0059}^{+0.0024}$ &
                        \\ $\Lambda_{\rm{\overline{MS}}}^{(f=5)}$, MeV
                        &  &
                         & $392_{-96}^{+44}$ &  \\ \hline
                        $\alpha_s(M_Z)$ ($N^4LO)$&  &  &   & $0.1277$
                        \\
                        $\Lambda_{\rm{\overline{MS}}}^{(f=5)}$, MeV &  &    &  & $345$ \\ \hline \hline

                        $\chi_1^2 (KEDR)/ndf$       &
                        5.433/20                     &
                        5.158/20               &
                        3.605/20&
                        3.851/20 \\
                        P($\chi^2$), $\%$ &
                        99.947     &
                        99.965     &
                        99.998     &
                        99.997    \\\hline \hline
                        $\Lambda_{\rm{\overline{MS}}}^{(f=3)}$, MeV
                        &
                        $372_{-166}^{+150}$   &
                        $413_{-194}^{+194}$     &
                        $668_{-193}^{+63} $&
                        $597$\\ \hline
                        $\nu$               &
                        $ 0.990 \pm 0.014$   &
                        $ 0.991 \pm 0.015$    &
                        $ 0.997 \pm 0.011$ &
                        $ 0.992$\\ \hline
                        $\alpha_s(m_\tau)$ &
                        $0.3178_{-0.0807}^{+0.0872}$   &
                        $0.3571_{-0.1011}^{+0.1313} $  &
                        $0.5748_{-0.1721}^{+0.0855} $ &
                        $0.478$\\      \hline
                        %\hline
                        $\alpha_s(M_Z)$ ($NLO)$   &
                        $0.1150_{-0.0110}^{+0.0075}$   &  &     &
                        \\
                        $\Lambda_{\rm{\overline{MS}}}^{(f=5)}$, MeV & $190_{-98}^{+98}$ &  &  &  \\ \hline
                        %\hline
                        $\alpha_s(M_Z)$ ($NNLO)$   &   &
                        $0.1187_{-0.0125}^{+0.0093} $ &  &\\
                        $\Lambda_{\rm{\overline{MS}}}^{(f=5)}$, MeV &  & $216_{-116}^{+132}$ &  &  \\ \hline
                        %\hline
                        $\alpha_s(M_Z)$ ($N^3LO)$ & &   &
                        $0.1303_{-0.0086}^{+0.0025} $&  \\
                        $\Lambda_{\rm{\overline{MS}}}^{(f=5)}$, MeV &
                        &
                        & $390_{-134}^{+46}$ &  \\ \hline
                        %\hline
                        $\alpha_s(M_Z)$ ($N^4LO)$&  &
                        & & $0.1273$ \\
                        $\Lambda_{\rm{\overline{MS}}}^{(f=5)}$, MeV &  &  &  & $339$
                        \\ \hline
                \end{tabular}

        \end{center}
\end{table}

The freedom in choosing   the model of  Eq.(\ref{Chi2:2}) for defining  the minimization function $\chi^2_1$  with the additional  free parameter $\nu$ of the fits  reflects the possibility
of taking into account additional systematical uncertainties in  comparison with the  massless perturbative QCD approximation of R(s) with the related   22  KEDR data points  fixed from  the results of Ref.\cite{KEDR:2018hhr} with their
 minimal experimental systematical  uncertainty $\sigma_0=0.0171$.

Considering carefully the results given in Table \ref{Tab:Results1},   we arrive at  the following conclusions :
\begin{itemize}
        \item the values of the characteristic functions $\chi_0^2$ and $\chi_1^2$  decrease  after taking into account explicitly evaluated higher order QCD corrections to the R ratio up to the level of the  $O(\alpha_s^4)$ contributions and  increase  after using the $O(\alpha_s^5)$ estimates;
        \item the  extracted NLO  values of  $\Lambda_{\rm{\overline{MS}}}^{(f=3)}$ , $\Lambda_{\rm{\overline{MS}}}^{(f=5)}$ and the
        related  ones for  $\alpha_s(m_{\tau})$  and $\alpha_s(M_Z)$  increase after taking into
                account the   NNLO and N$^3$LO QCD contributions to the  R ratio in the fits  ;
                \item the difference between  the NLO $\chi_0^2$ related results  of  Table \ref{Tab:Results1} and the ones of Ref.\cite{KEDR:2018hhr}  $\Lambda_{\rm{\overline{MS}}}^{(f=3)}=361^{+155}_{-174}$ MeV and $\alpha_s(m_{\tau})=0.332^{+0.100}_{-0.092}$ are in part due to the fact that
        instead  of the  NLO inverse log $\alpha_s$  approximation of Eq.(\ref{NLO}) the NNLO of Eq.(\ref{NNLO}) was   used;
                \item the introduction of the extra fitting parameter $\nu$ into the definition of the minimized characteristic  function $\chi_1^2$  leads  to a  definite
                decrease in  the extracted   QCD coupling constant values   $\alpha_s(m_{\tau})$ and $\alpha_s(M_Z)$ ;
                \item the estimate of systematic uncertainties of the fits  fixed by the difference of the results obtained in the cases of consideration of two minimization
                function models  $\chi_0^2$ and $\chi_1^2$ are smaller than the errors  defined by summing in quadrature  the statistical and systematical uncertainties of R ratio data  obtained by KEDR ;
                \item the $\chi_1^2$ model motivated NLO $\alpha_s(M_Z)$ value is in good agreement with  the Particles Data Group  world average value of $\alpha_s(M_Z)$;
                \item the NLO and  NNLO $\chi_1^2$ results
                $\alpha_s(M_Z)=0.1150_{-0.0110}^{+0.0075}$  and
                $\alpha_s(M_Z)=0.1187_{-0.0125}^{+0.0093}$ are in good agreement with other extractions of $\alpha_s(M_Z)$ from  R ratio data in higher energy regions
                 (see Refs.\cite{Kuhn:2001dm,Zenin:2001yx}) ;
                \item  the  N$^3$LO  perturbative QCD  contributions
                  to the  R ratio   move  the
                   $\alpha_s(M_Z)=0.1303_{-0.0086}^{+0.0025}$ value towards   higher values, not supported by  determination from other processes ;
                \item the inclusion into  the fits of the estimated N$^4$LO contributions result in a certain  decrease of the
        N$^3$LO outcomes  ;
\end{itemize}

In general,  our considerations of KEDR data are  in agreement with  the   results of the fits
of  Ref.\cite{Boito:2025qfz} and  feel  theoretical uncertainties
better than the ones given  in Ref.\cite{Shen:2023qgz} .

 \section{Results of fits of BESIII data}

 It is known  that  there is a  definite  tension between the
 N$^3$LO QCD approximation for R(s)  obtained in Ref.\cite{Baikov:2008jh}
   and the experimental results of BESIII collaboration \cite{BESIII:2021wib}.
  This problem was already raised in the original experimental work,  where it was emphasized that in the {\bf energy range between 3.4 GeV and 3.6 GeV
         eight   BESIII data points for R lie  above
   the corresponding experimental  KEDR results } though at  still not disturbing  1.9 $\sigma$  standard deviations level  and are higher at the 2.9 $\sigma$  standard deviations level than the theoretical   N$^3$LO  QCD predictions   for R(s) provided $\alpha_s$  is  fixed  as
  the world average value $\alpha_s(M_Z)\approx 0.118$. Moreover,  it was recently emphasized  in Ref. \cite{Boito:2025qfz} that the  tension between BESIII results and perturbative QCD expressions  remains at more that 3$\sigma$ level  after taking into account  the remaining 6 BESIII data points
  below 3.4 GeV to 2.23 GeV.

  In this work,  we have  made a  further step  in considering  the original BESIII data .  In  Table \ref{Tab:BESdata}  we  summarized the  re-presentation   all 14 BESIII data points
  with taking explicitly
  into account    systematical uncertainties, only commented in the original work of Ref.\cite{BESIII:2021wib}.
  The outcomes of the determinations of
  $\Lambda_{\rm{\overline{MS}}}^{(3)}$ and  $\alpha_s$  values are  presented in Tables  \ref{Tab:Results3}, \ref{Tab:Results4},   below.
   In  the latter case,  the normalization systematical uncertainty factor $\sigma_0=0.0349$  in the denominator of Eq.(28) is fixed by the minimal entry to the last column of Table 1.
 \begin{table}[H]
        \begin{center}
                \caption {\label{Tab:Results2}
                        Results of fitting all 14  BESIII data points.
                }
                \begin{tabular}[c]{|c|c|c|c|c|c|} \hline R(s)  approximation
                        & $O(\alpha_s^2)$ & $O(\alpha_s^3)$ & $O(\alpha_s^4)$ &$O(\alpha_s^5)$ estimate \\ \hline  \hline
                        $\chi_0^2 (BESIII)/ndf$        &  53.723/13    &  53.603/13          & 53.484/13  & 53.386/13 \\
                         $P(\chi^2)$, $\%$   &   $6.76\times 10^{-5}$
                        &  $7.10\times 10^{-5}$         &  $7.44\times 10^{-5}$  &$7.74\times 10^{-5}$ \\\hline \hline

                        $\Lambda_{\rm{\overline{MS}}}^{(f=3)}$, MeV     & $184_{-63}^{+65}$   & $607_{-66}^{+71}$  & $573_{-78}^{+92}$&        $229$ \\ \hline\hline
                        $\chi_1^2 (BESIII)/ndf$                      &  51.183/12    &  51.176/12          & 51.166/12 & 53.386/12 \\
                            $P(\chi^2)$, $\%$   &     $8.65\times 10^{-5}$
                        &     $8.67\times 10^{-5}$         &     $8.71\times
                        10^{-5}$   &    $7.96\times 10^{-5}$ \\\hline \hline
                        $\Lambda_{\rm{\overline{MS}}}^{(f=3)}$, MeV     & $19_{-18}^{+102}$   & $18_{-17}^{+104}$  &
                        $20_{-19}^{+118}$&          $20$ \\\hline
                        $\nu$    & $0.96\pm 0.02$  & $0.96\pm 0.02$&          $0.96\pm 0.02$   & $0.96$  \\        \hline
                \end{tabular}
\end{center}
\end{table}
        \begin{table}[H]
        \begin{center}
                \caption {\label{Tab:Results3}
                        Results of fitting the truncated   BESIII
                data with points  below the mass of $J/\Psi$ meson.
                }
                \begin{tabular}[c]{|c|c|c|c|c|c|} \hline R(s) approximation
                        & $O(\alpha_s^2)$ & $O(\alpha_s^3)$ & $O(\alpha_s^4)$ & $O(\alpha_s^5)$ estimate \\ \hline  \hline
                                $\chi_0^2 (BESIII)/ndf$ (6 points)
                                &  8.128/5         &  7.377/5   &
                                12.596/5  & 17.778/5 \\
                                   $P(\chi^2)$, $\%$    &     14.932  &     19.941   &    2.747  &    0.324 \\\hline \hline

                        $\Lambda_{\rm{\overline{MS}}}^{(f=3)}$, MeV     &  $527_{-67}^{+59}$   & $607_{-89}^{+84}$  & $573_{-86}^{+67}$&          $477$ \\ \hline
                        $\alpha_s(m_\tau)$  &   $0.4087_{-0.0418}^{+0.0425}$   &
                        $0.488_{-0.0665}^{+0.0781}$  &
                        $0.478_{-0.0678}^{+0.0654} $ & $0.3938$\\
                        \hline
                          $\alpha_s(M_Z)$ ($NLO)$&   $0.1227_{-0.0031}^{+0.0026}$ &                      & & \\
                        $\Lambda_{\rm{\overline{MS}}}^{(f=5)}$, MeV &     $292_{-45}^{+41}$ &  &  &  \\ \hline
                        $\alpha_s(M_Z)$ ($NNLO)$&            &    $0.1280_{-0.0040}^{+0.0035}$                  &  &\\
                        $\Lambda_{\rm{\overline{MS}}}^{(f=5)}$, MeV &  &    $348_{-63}^{+61}$  &  &  \\ \hline
                        $\alpha_s(M_Z)$ ($N^3LO)$&                    &        &    $0.1263_{-0.0040}^{+0.0029} $ &  \\
                        $\Lambda_{\rm{\overline{MS}}}^{(f=5)}$, MeV &  &  &    $322_{-60}^{+48}$
                        &  \\ \hline
                        $\alpha_s(M_Z)$ ($N^4LO)$&  &                                 & &    $0.1218$ \\
                        $\Lambda_{\rm{\overline{MS}}}^{(f=5)}$, MeV &  &  &  &    $257$ \\ \hline \hline
                        $\chi_1^2 (BESIII)/ndf$ (6 points)      &
                        6.011/4          &  5.798/4   & 5.500/4  &
                        6.112/4 \\\hline \hline
                        $P(\chi^2)$, $\%$    &  19.832  &  21.148   & 23.972 & 19.094 \\\hline \hline

                        $\Lambda_{\rm{\overline{MS}}}^{(f=3)}$, MeV     &
                        $375_{-156}^{+130}$   & $424_{-188}^{+173}$  &
                        $566_{-262}^{+111}$&  $463$ \\ \hline
                        $\nu$              &   $0.966 \pm 0.023$   & $0.969 \pm
                        0.024$  &    $0.965 \pm 0.013$&        $0.955$ \\\hline
                        $\alpha_s(m_\tau)$ &   $0.3194_{-0.0762}^{+0.0746}$   &   $0.3638_{-0.0994}^{+0.1163} $
                        &$0.4723_{-0.1716}^{+0.1136} $& $0.3849$\\ \hline
                        $\alpha_s(M_Z)$ ($NLO)$   &   $0.1151_{-0.0101}^{+0.0065}$     &   &   &\\
                        $\Lambda_{\rm{\overline{MS}}}^{(f=5)}$, MeV &    $192_{-93}^{+84}$  &  &  &  \\ \hline
                        $\alpha_s(M_Z)$ ($NNLO)$  &  &   $0.1193_{-0.0118}^{+0.0082}$    & &  \\
                        $\Lambda_{\rm{\overline{MS}}}^{(f=5)}$, MeV &  &    $224_{-114}^{+118}$  &  & \\ \hline
                        $\alpha_s(M_Z)$ ($N^3LO)$ &   &   &        $0.1260_{-0.0139}^{+0.0047}$ &  \\
                        $\Lambda_{\rm{\overline{MS}}}^{(f=5)}$, MeV &  &  &    $317_{-170}^{+79}$ &   \\ \hline
                        $\alpha_s(M_Z)$ ($N^4LO)$ &   &   &      &    $0.1211$ \\
                        $\Lambda_{\rm{\overline{MS}}}^{(f=5)}$, MeV &  &  &  &    $247$ \\ \hline
        %       \begin{tabular}[c]{|c|c|c|c|c|c|} \hline

                \end{tabular}
        \end{center}
\end{table}

From the results given in Table \ref{Tab:Results2} it  is clear that both ways of fitting the whole set of available  BESIII data produce  unsatisfactory results.
Indeed, in the first case,  in addition to an  uncomfortably large value of the minimization function $\chi_0^2 \approx 50/13$ (?!) (compare with the  presented in Table 2 above values , which vary between  $\chi_0^2 \approx 6/21$
to $\chi_0^2 \approx 4/21$ , achieved in our  analysis of KEDR data ) the unclear variation of the values of $\Lambda_{\rm{\overline{MS}}}^{(3)}$ from 150 MeV  to 600 MeV
with returning to over 230 MeV again were detected. Moreover, the
attempt to take into account the possible extra  systematical uncertainties in the fits of BESIII data in addition to the problem
 of the appearance of high value of over
$\chi_1^2 \approx 51/12$  (?!)   leads us to
the manifestation of the problem of  a non-physical    small  value of the parameter $\Lambda_{\rm{\overline{MS}}}^{(3)}\approx 20$ MeV (?!).

To understand whether it is possible to get from BESIII any physical information on the problems of "measuring" the value of the strong interactions coupling constant $\alpha_s$ we excluded from the analysis
8 data points above the mass of $J/\Psi$ meson in the
energy region between 3.4 GeV to 3.8 GeV. As was already noticed
by the members of the  BESIII collaboration, these 8 points are  responsible  for the certain  tensions between the data sets of the  BESIII and KEDR collaborations \cite{BESIII:2021wib}. The results of the fits
of these truncated  BESIII data are  presented in Table 4 and demonstrate
 agreement with the  results of the  fits of KEDR
 data presented in Table 2
 within the  existing specified concrete uncertainties of both data sets and the performed fits.

\section{Combined   fits of KEDR and  truncated BESIII data}

To get now more statistically substantiated results,  we perform the  combined fits of    22 data points from KEDR data and  6 data points of the truncated BESIII data.
The adjustment of two sets of data is made by introducing two models for  $\chi_1^2$ parameter for KEDR and BESIII data sets with
two
extra free parameters $\nu$ and $\nu_1$, responsible for fixing systematical uncertainties of the  KEDR and BESIII data sets.  A similar  procedure was previously
applied in the combined NLO fits of the available $e^+e^-$ data above the production  thresholds  of charm and bottom quarks in Ref.\cite{Marshall:1989yi}.
The results are presented in Table \ref{Tab:Results4} below.
\begin{table}[H]
        \begin{center}
                \caption {\label{Tab:Results4}  Outcomes  of fitting
                  truncated  BESIII and KEDR data.  }
\begin{tabular}[c]{|c|c|c|c|c|c|} \hline R(s)  approximation
        & $O(\alpha_s^2)$ & $O(\alpha_s^3)$ & $O(\alpha_s^4)$ & $O(\alpha_s^5)$ estimate \\ \hline  \hline
        $(\chi_1^2(BESIII)+\chi_1^2(KEDR))/ndf$         &
        11.445/25&
        10.958/25 &
        9.423/25&
        10.419/25  \\
            P($\chi^2$), $\%$         &
        99.050&
        99.315 &
        99.794 &
        99.536  \\\hline \hline
        $\Lambda_{\rm{\overline{MS}}}^{(f=3)}$, MeV      &
        $ 374_{-113}^{+101}$  &
        $ 420_{-136}^{+131}$   &
        $ 618_{-162}^{+68}$   &
        $ 518$\\ \hline
        $\nu$ (BES)         &
        $0.966 \pm  0.019$  &
        $0.969   \pm  0.020  $
        & $ 0.964\pm  0.013 $
        & $0.955$     \\ \hline
        $\nu$ (KEDR)        &
        $0.990 \pm  0.012$   &
        $0.992 \pm 0.013$   &
        $0.996 \pm 0.013$  &
        $0.990$\\ \hline
        $\alpha_s(m_\tau)$  &
        $0.3187_{-0.0556}^{+0.0564}$  &
        $0.3607_{-0.0725}^{+0.0836}$  &
        $0.5201_{-0.1305}^{+0.0769} $ &
        $0.4203$\\ \hline
        $\alpha_s(M_Z)$ ($NLO)$&$0.1151_{-0.0069}^{+0.0052}$ &                      & & \\
        $\Lambda_{\rm{\overline{MS}}}^{(f=5)}$, MeV &$191_{-68}^{+65}$ &  &  &  \\
       \hline
        $\alpha_s(M_Z)$ ($NNLO)$&     &$0.1190_{-0.0081}^{+0.0064}$                  &
       &\\
        $\Lambda_{\rm{\overline{MS}}}^{(f=5)}$, MeV &  & $220_{-84}^{+88}$  &  &  \\ \hline
        $\alpha_s(M_Z)$ ($N^3LO)$&                                 &
        & $0.1283_{-0.0075}^{+0.0028} $ &  \\
        $\Lambda_{\rm{\overline{MS}}}^{(f=5)}$, MeV &  &  & $354_{-111}^{+49}$
        &  \\ \hline
        $\alpha_s(M_Z)$ ($N^4LO)$&  &                                 & & $0.1238$ \\
        $\Lambda_{\rm{\overline{MS}}}^{(f=5)}$, MeV &  &  &  & $284$ \\ \hline
     \end{tabular}
\end{center}
\end{table}
The results of the combined fits of 22 points of the  KEDR data set   and 6  points of the  BESIII  data set are illustrated by the plots of Figures 1,  where  all shifted points of the  KEDR and BESIII collaborations given in Ref. \cite{KEDR:2018hhr} and Ref.\cite{BESIII:2021wib}, are represented as well. The shifts are fixed after the proper adjustments by    multiplying  them by  the related  values of the  factors  $\nu (KEDR)$ and $\nu (BESIII)$ from Table 5  obtained in the process of the  NLO fits. The theoretical curves at these plots  are drawn after the substitution of the
related fitted outcomes  for the range of   $\Lambda_{\rm{\overline{MS}}}^{(3)}$ values into the $O(\alpha_s^2)$ and
$O(\alpha_s^3)$  approximations for R(s) from Eq.(20)
with the truncated inverse logarithmic approximations for $\alpha_s(s)$,
defined at the NLO, NNLO level by Eq.(8) and Eq.(9) respectively.

Considering the curves of  Figure 1,  we arrive at the following
conclusion :
\begin{itemize}
        \item  in the cases of the  obtained NLO and NNLO range of
        values
          $\Lambda_{\rm{\overline{MS}}}^{(3)}=374_{-113}^{+101}$ MeV and
          $\Lambda_{\rm{\overline{MS}}}^{(3)}=420_{-136}^{+131}$ MeV ,
        which correspond to the given in Table \ref{Tab:Results4}
         results  for $\alpha_s(M_Z)$,  respected by the analysis
        of other experimental data (see the considerations in the reviews of Refs.\cite{Salam:2017qdl,Pich:2020gzz,dEnterria:2022hzv}    ) all experimental data points ,
        apart of 6  truncated  BESIII points in the energy region above 3.4 GeV, are  in agreement with the results of the  NLO and NNLO  perturbative QCD analysis.
        \end{itemize}

        Concerning the deviations of the  6 BESIII data points from the depicted curves,  we would like to make the following observations :
        \begin{itemize}
                \item  these points lie in the energy  region between the masses of the  $J/\Psi(3096)$ and $\Psi'(3770)$  mesons, which,  due to
                considerations of the  KEDR collaboration, are   still well described by the opened production of  (anti)quark pairs  with $f=3$ numbers of light flavours only ;
                \item these points belong to the   energies of production of  lepton $\tau^+\tau^-$  pairs, which can  decay further on into light hadrons;
        \end{itemize}

        Despite the fact that these physical effects seem to be taken into
        account in  the discussed above experimental analysis of both the  KEDR and BESIII collaborations, it may be of interest to further consider  the problem  of  comparison  between the points of these data sets in future.

%\newpage

\begin{figure*}
        \begin{center}
                \includegraphics[width=0.5\textwidth]{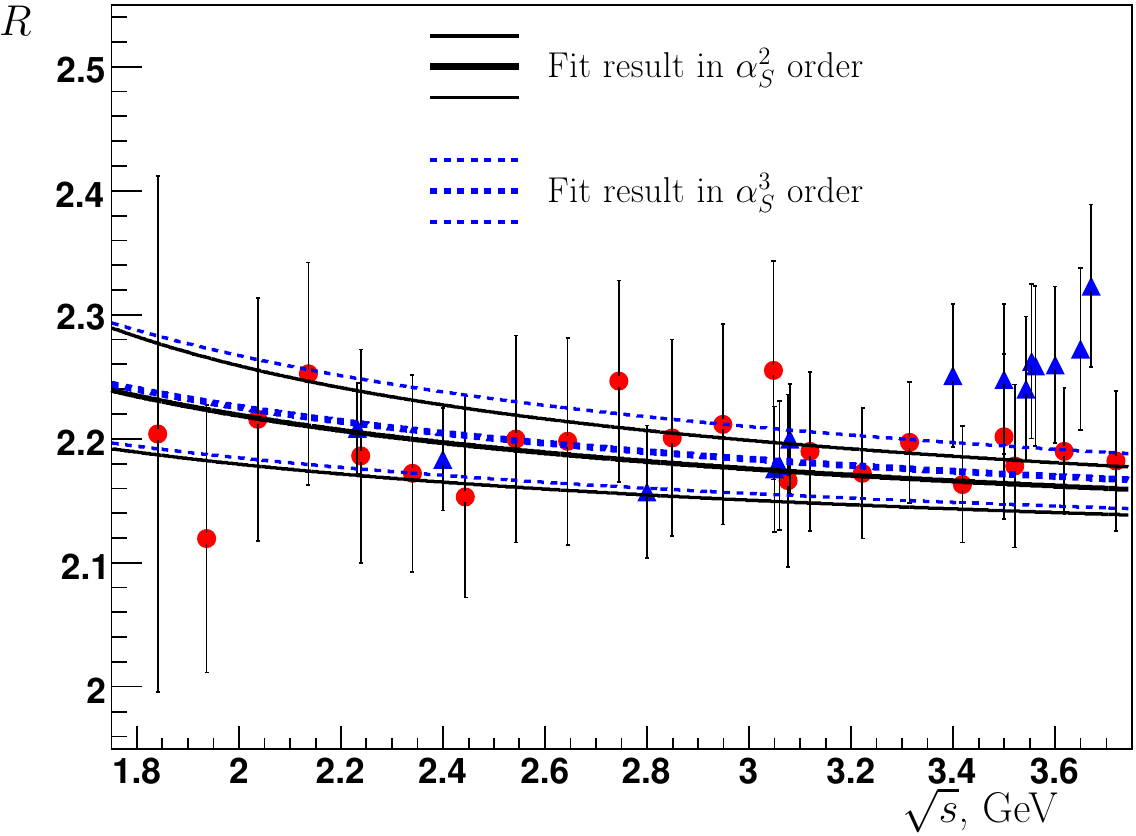}
               \caption{\label{Fig:CombinedFit23} The result of the
                 joint fit of the KEDR (marked with circles) 
                 and truncated BESIII (marked with triangles) 
                 experimental data. The thin lines correspond to the range specified by  $\Lambda$
                 parameter errors.
                 The experimental data are presented taking into account the
                 coefficients $\nu$ obtained from fitting in
                 $\alpha_s^2$ order.
                 }
                        \end{center}
 \end{figure*}

  \begin{figure*}
        \begin{center}
               \includegraphics[width=0.5\textwidth]{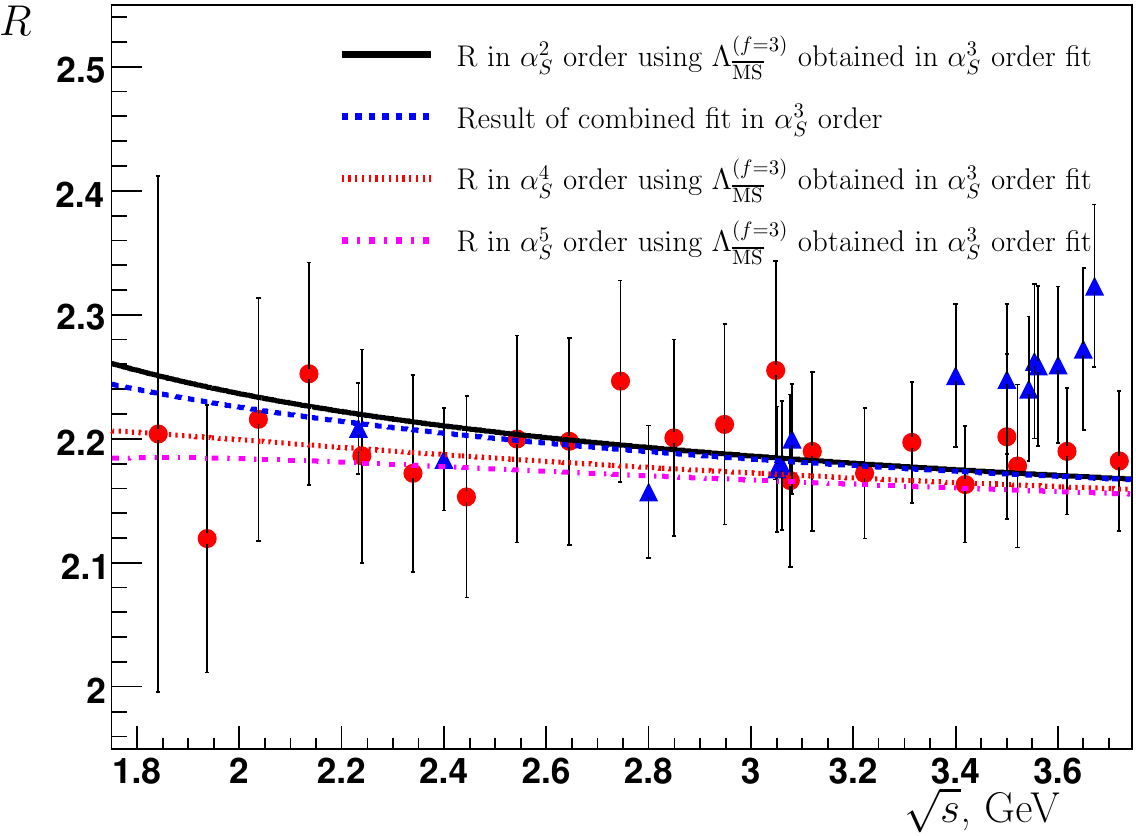}
                \caption{\label{Fig:CombinedFit2345} R(s)  curves obtained in different
                        orders of perturbative QCD with  $\Lambda_{\rm{\overline{MS}}}^{(3)}$  fixed by  the result of
                        joint fit  of KEDR (marked with circles)   and
                        the truncated   BESIII (marked with triangles) 
                        data.
                       The experimental data are presented taking into account the
                        coefficients $\nu$ obtained from fitting in $\alpha_s^2$ order.}

\end{center}
\end{figure*}

\vspace{3cm}

In Figure 2, we present the related plots for R(s) drawn for the
truncated
at the NLO, NNLO, NNLO and N$^3$LO terms theoretical expressions for R from
Eq. (20) with the QCD coupling constant.  The value of the QCD coupling
constant $\alpha_s(s)$ is determined at the NNLO order through the inverse
log approximation of Eq. (9) with the value of the parameter $\Lambda_{\rm{\overline{MS}}}^{(3)}=420$ MeV, obtained in the process of our combined NNLO fits .

Two higher lines correspond to the  NLO and  NNLO $O(\alpha_s^2)$ and $O(\alpha_s^3)$ approximations for the  R ratio.
Two lower   ones correspond to the  not fitted    $O(\alpha_s^4)$ and $O(\alpha_s^5)$   related theoretical approximants  for R(s).

Notice that in the whole considered region of energies considered the  N$^3$LO
perturbative QCD corrections are  larger than the NNLO ones. This
happens due to the manifestation of the following pattern in general
{\bf not asymptotic in the Minkowskain region} fixed order
perturbative QCD expression  for R(s) :

\begin{equation}
        \label{expand}
        R(s)
        = 2\bigg[1+a_s+\underline{1.6398}a_s^2+ \big(\underline{6.3710}-\underline{\underline{16.6550}}\big)a_s^3+\big(\underline{49.0757}-\underline{\underline{155.9555}}\big)a_s^4 \\ \nonumber
        +\big(\underline{\sim 275}-\underline{\underline{780}})a_s^5+O(a_s^6)\bigg]~~.
\end{equation}

Here  the  underlined coefficients are the values of the
    directly in the Euclidean region  $\rm{\overline{MS}}$-scheme  terms, which result from the evaluation of  the concrete sets of multiloop Feynman diagrams,  and the additional terms with  double lines are the contributions of the  terms shown in  Eqs.(17) , which
appear due to  the analytical continuation  from  the Euclidean to the Minkowski region.   They are  modifying  the possibly
{\bf asymptotic n! growth} of the related   $\rm{\overline{MS}}$ -scheme  coefficients of fixed order perturbative QCD  expansions  in the Euclidean region.

These additional  contributions  are responsible for
the change of the  structure of the perturbative QCD expansion from
 sign constant asymptotic series  in the Euclidean region to the
  non-regular in sign power expansion, which in case of choosing as the value of
   $\Lambda_{\rm{\overline{MS}}}^{(3)}=420$ MeV and thus    $\alpha_s(m_{\tau})$=
0.361 from the presented in Table \ref{Tab:Results4} results lead to the  following numerical behaviour of the
considered in Eq.(29)  perturbative QCD approximation:

\begin{equation}
        R(s)|_{s=m_{\tau}^2} =2\bigg[1+0.1149+0.0217-0.0156-0.0186-0.0101+\dots\bigg]\\
\end{equation}

The appearance of  not small negative contributions in the expressions for R(s) are related with the negative contributions to the coefficients of R(s) of sizable kinematical $\pi^2$ effects and are responsible for the appearance of the  observed by us  tendency  of  growth of the
extracted values of the QCD coupling constant $\alpha_s$ in the
process of applications of the truncated, at different levels,  fixed
order  perturbation theory (FOPT)  QCD approximations for R(s) in the
energy region with $f=3$ numbers of active flavours.

To be more definite,  in the final part of this work we present the  results  based on the combined fits of  KEDR and the  truncated BESIII experimental data

\begin{eqnarray}
        ~~~~~~~~~~~NLO~~~\alpha_s(M_Z)&=&0.1151_{-0.0069}^{+0.0052}~~~  \\
        ~~~~~~~NNLO~~\alpha_s(M_Z)&=&0.1190_{-0.0081}^{+0.0064}~~~~~ \\
        ~~~~~~~N^3LO~~\alpha_s(M_Z)&=&0.1283_{-0.0075}^{+0.0028}~~~~ \\
         ~~~~~~N^4LO~~~\alpha_s(M_Z)&=&\sim 0.1238 ~~~~~~~~~
        \end{eqnarray}
        
  which are closed to the following similar   similar results of the fits of KEDR data alone 
        
\newpage

\begin{eqnarray}
	~~~~~~~~~~~NLO~~~\alpha_s(M_Z)&=&0.1150_{-0.0110}^{+0.0075}~~~  \\
	~~~~~~~NNLO~~\alpha_s(M_Z)&=&0.1187_{-0.0125}^{+0.0093}~~~~~ \\
	~~~~~~~N^3LO~~\alpha_s(M_Z)&=&0.1303_{-0.0081}^{+0.0025}~~~~ \\
	~~~~~~N^4LO~~~\alpha_s(M_Z)&=&\sim 0.1273 ~~~~~~~~~.
\end{eqnarray}

Note that we consider the agreement of our NNLO result with the
ones obtained in the process of the performed in
 Ref.\cite{Shen:2023qgz} comparisons
 of   KEDR data  with the N$^3$LO  approximations for R, fixed by applying
   the proposed in Ref.\cite{Brodsky:2013vpa} procedure of dealing with   high order
 perturbative QCD expansions,  as rather accidental one. Indeed, the application of this procedure
 to the series for R(s) defined by Eq.(\ref{expand})
 presumes further expansion of  underlined contributions to the
 coefficients of perturbative series of Eq.(\ref{expand}) to scale-independent and scale non-independent parts  and  further
 absorption  of the latter ones,  together with double underlined analytical coefficient ,  into the properly defined  scales of $\alpha_s$  in   Eq.(\ref{expand}).
 However, it is known from the considerations of Refs.\cite{Kataev:2023xqr,Mikhailov:2024mrs} that the analysis  of Refs.\cite{Shen:2023qgz,Brodsky:2013vpa}
 contain the number of still non-fixed ambiguities, which may be fixed in the process of possible further studies .
 Other  interesting issues may be the further applications of the low-energy $e^+e^-$ annihilation to hadrons data
 for the analysis of the number of  new theoretical problems, discussed in \cite{Boito:2024gtb,Pich:2022tca}
  and quite recenty in \cite{Beneke:2025hlg}
 in the case of semi hadronic decay width of  $\tau$-lepton.

\section{Conclusions}

We have  demonstrated  that the   direct fits  of   KEDR and BESIII  collaboration data for the
$e^+e^-$ hadrons  R  ratio  below the thresholds of production of charm quarks by  the order-by-order truncated
 perturbation theory QCD expressions  for the  R ratio
and the QCD $\beta$-function allow one  to get
new useful information   on the values of $\alpha_s(M_Z)$ provided
BESIII data are
is truncated  from above by the energy scale of the
 the mass of J/$\Psi$ meson. The  new values of  $\alpha_s(M_Z)$,
 extracted by us while applying  the   NLO and NNLO approximations ,
 are in agreement with other  determinations of  $\alpha_s$, including the ones from  higher energy    $e^+e^-$ annihilation to hadrons data .
 A certain  tendency of  growth
 of $\alpha_s(M_Z)$ values starting from NLO to N$^3$LO orders
 of applied perturbative expansions is  detected.
 Further continuation of the related studies may be useful in clarifying whether it is possible to get more solid information on the
value  of  $\alpha_s(M_Z)$ from further analysis of the $e^+e^-$ annihilation to hadrons R(s) data in the regions below and  above the thresholds of production of pairs of charmed quarks for
clarifying the possibility of the manifestations  of the residual quark-hadron  duality-violating effects Ref.\cite{Shifman:2000jv}  considered from various points of view in the
number of works on the subject \cite{Boito:2025qfz, Boito:2024gtb,Pich:2022tca} and of
  careful treatment of high order perturbative  contributions with the help of  CIPT or APT   related  QCD considerations.

 The results of taking into account of the explicitly unknown , but    estimated ,  $\alpha_s^5$ (N$^4$LO)  QCD
contributions may be further used to  simulate  in part  theoretical uncertainties of
the N$^3$LO studies, which within other terms may be more definitely specified  by the considerations of the theoretical  studies 
of scale-scheme uncertainties, considered in the detailed reviews of Refs.\cite{Deur:2016tte, Salam:2017qdl, Pich:2020gzz} and 
studied in various   massless analysis of  QCD predictions for R(s) in Ref.\cite{Shen:2023qgz} (with the discussed in Refs.\cite{Kataev:2023xqr,
Mikhailov:2024mrs} limitations 
) and in Refs. \cite{Kazakov:1985qh,Chyla:1991ca,Stevenson:2012ti}
previously.  We hope that following these considerations at the $\alpha_s^5$-level will not change the 
considered by us tendency of the non-regular behaviour of the  $\overline{MS}$-scheme    $O(\alpha_s^4)$  QCD approximants in the region of energies 
below charm quark-anti quark  production.  This feeling may be either supported or restricted by  more careful analysis of this problem in future with taking into  the dependence from masses of  quarks as well.

\section{Acknowledgements}

The preliminary results of this work were reported at a   scientific Seminar at  INR RAS Theory Division 17.11.25.
We would like to express our  gratitude  to
Konstantin  Chetyrkin and in particular  to   Nikolai  Krasnikov
for   supporting conversations. We also wish to thank  Andrej  Arbuzov, Dmitry  Gorbunov, Ivan Logashenko and Sergey Mikhailov for productive discussions.

The material of this work was further on presented at  Quarks-2026 International Seminar, Perozavodsk, 19.05.26 . We would like to thank Diogo Boito, Tatyana Kharlamova 
and Oleg Zenin for  useful comments. 
\section{Appendix}
The calculations performed in this work were carried out in the approximation of massless quarks.
The influence of the s-quark mass can be estimated
from the expansions given in Ref. \cite{Gorishnii:1986pz,Chetyrkin:1997qi},   which in QED limit should be compared with the obtained in Ref.\cite{Onishchenko:2024pcz} 
 explicit analytical results.

The results obtained are shown in Table \ref{Tab:Results5}.

\begin{table}[H]
        \begin{center}
                \caption {\label{Tab:Results5}  Results of fitting
                  truncated BESIII and KEDR data when considering a
                  non-zero s-quark mass
                  }.
\begin{tabular}[c]{|c|c|c|c|} \hline R(s)  approximation
        & $O(\alpha_s^2)$ & $O(\alpha_s^3)$  \\ \hline  \hline
        $(\chi_1^2(BESIII)+\chi_1^2(KEDR))/ndf$ &  11.590/25&    11.274/25         \\
        ($\hat{m}_s=0.25$~GeV)  & &  \\        \hline
        $P(\chi^2)$                       & 98.957 & 99.151                  \\\hline
        $\Lambda_{\rm{\overline{MS}}}^{(f=3)}$, MeV      &   $361_{-110}^{+99}$  &  $ 387_{-125}^{+117}$         \\ \hline
        $\nu$ (BES)         &          $0.966 \pm  0.019$  &          $0.967 \pm  0.019$     \\ \hline
       $\nu$ (KEDR)        &         $0.990 \pm  0.012$  &          $0.991 \pm  0.013$  \\ \hline
        $\alpha_s(m_\tau)$  &  $0.3124_{-0.0539}^{+0.0543}$  &$0.3423_{-0.0651}^{+0.0700}$                \\ \hline
        $\alpha_s(M_Z)$ ($NLO)$&$0.1144_{-0.0070}^{+0.0052}$ &                       \\
        $\Lambda_{\rm{\overline{MS}}}^{(f=5)}$, MeV &        $183_{-66}^{+63}$ &    \\ \hline
        $\alpha_s(M_Z)$ ($NNLO)$&     &$0.1172_{-0.0078}^{+0.0061}$                  \\
        $\Lambda_{\rm{\overline{MS}}}^{(f=5)}$, MeV &  & $199_{-76}^{+77}$    \\ \hline
     \end{tabular}

\end{center}
\end{table}

It can be concluded that the non-zero mass of the strange quark (we are using the given e.g. in Ref. \cite{Gorishnii:1986pz}  definition of the invariant strange quark mass $\hat{m}_s$)
   has a
noticeable effect on the calculation results, and the corresponding
change in the value of $\alpha_s(M_Z)$  is approximately $-2 \times
10^{-3}$.
However, the changes in the results are significantly smaller than
the errors in the presented values.

\end{document}